\documentclass[a4paper,10pt]{article}
\usepackage{mathtools}
\usepackage{mathrsfs} 
\usepackage[colorlinks=true, urlcolor=blue, linkcolor=red]{hyperref}
\usepackage{soul}
\usepackage{yfonts}

\usepackage{dsfont}
\usepackage{tikz-cd}

\RequirePackage{tikz-cd}
\RequirePackage{amssymb}
\usetikzlibrary{calc}
\usetikzlibrary{decorations.pathmorphing}
\usepackage{blindtext}
\tikzset{curve/.style={settings={#1},to path={(\tikztostart)
    .. controls ($(\tikztostart)!\pv{pos}!(\tikztotarget)!\pv{height}!270:(\tikztotarget)$)
    and ($(\tikztostart)!1-\pv{pos}!(\tikztotarget)!\pv{height}!270:(\tikztotarget)$)
    .. (\tikztotarget)\tikztonodes}},
    settings/.code={\tikzset{quiver/.cd,#1}
        \def\pv##1{\pgfkeysvalueof{/tikz/quiver/##1}}},
    quiver/.cd,pos/.initial=0.35,height/.initial=0}
\def\ba{\begin{equation}\begin{aligned}}
\def\ea{\end{aligned}\end{equation}}
\tikzset{tail reversed/.code={\pgfsetarrowsstart{tikzcd to}}}
\tikzset{2tail/.code={\pgfsetarrowsstart{Implies[reversed]}}}
\tikzset{2tail reversed/.code={\pgfsetarrowsstart{Implies}}}
\tikzset{no body/.style={/tikz/dash pattern=on 0 off 1mm}}
\usepackage{amsmath,tikz-cd}
\usepackage[english]{babel}
\usepackage[utf8x]{inputenc}
\usepackage{blindtext}
\usepackage[T1]{fontenc}
\usepackage[T2A]{fontenc}
\usepackage[a4paper,top=1.5 cm,bottom=2cm,left=1cm,right=1cm,marginparwidth=1.75cm]{geometry}
\usepackage{amsmath, amssymb}
\usepackage{yfonts}

\usepackage{mathtools}

\numberwithin{equation}{section}
\usepackage{wrapfig}
\usepackage{graphicx}
\usepackage[T1]{fontenc}
\usepackage{tikz}
\usetikzlibrary{arrows}
\usetikzlibrary{decorations.pathreplacing}
\usepackage[T2A]{fontenc}
\usepackage{color}
\usepackage{circuitikz} 
\graphicspath{{pic/}}
\definecolor{water} {rgb} {0.667, 0.855, 1}
\usepackage{pgfplots}
\usepackage{pgfplotstable}
\usetikzlibrary{circuits}
\usetikzlibrary{circuits.ee}
\usetikzlibrary{circuits.ee.IEC}
\usetikzlibrary{circuits.logic.IEC}
\usetikzlibrary{intersections}
\usepackage{tikz-cd} 
\usepackage{floatrow,graphicx,calc}
\usepackage{wrapfig}
\usepackage{yfonts}
\usepackage[natbib=true, backend = bibtex, style=numeric-comp, sorting=none,maxbibnames=99]{biblatex}
\usepackage{authblk}
\newtheorem{Def}{Def.}[section]
\newtheorem{Th}{Th.}[section]
\newtheorem{conjecture}{Conjecture}[section]
\usepackage{graphicx}
\def\sl{\mathfrak{sl}}
\def\KZ{{\rm KZ}}
\def\CR{{\cal R}}
\def\id{{\rm id}}

\date{}

\begin{document}

\title{\bf On connection between associators and 6j-symbols}

\author[1,2,3]{{\bf E. Lanina}\thanks{\href{mailto:lanina.en@phystech.edu}{lanina.en@phystech.edu}}}
\author[1]{{\bf A. Novokhatnii}\thanks{\href{mailto:novokhatnii.aa@phystech.edu}{novokhatnii.aa@phystech.edu}}}
\author[1,2,3]{{\bf A. Sleptsov}\thanks{\href{mailto:sleptsov@itep.ru}{sleptsov@itep.ru}}}
\author[1]{{\bf R. Stepanov}\thanks{\href{mailto:stepanov.rt@phystech.edu}{stepanov.rt@phystech.edu}}}

\vspace{5cm}

\affil[1]{Moscow Institute of Physics and Technology, 141700, Dolgoprudny, Russia}
\affil[2]{Institute for Information Transmission Problems, 127051, Moscow, Russia}
\affil[3]{NRC "Kurchatov Institute", 123182, Moscow, Russia}
\renewcommand\Affilfont{\itshape\small}

\maketitle

\vspace{-7cm}

\begin{center}
	\hfill MIPT/TH-28/26\\
	\hfill ITEP/TH-37/26\\
	\hfill IITP/TH-31/26
\end{center}

\vspace{4.7cm}

\begin{abstract}

We propose a new formula relating the rationalized Drinfeld associator obtained from the Knizhnik--Zamolodchikov equation to the quantum \(6j\)-symbols of the algebra \(\mathcal{U}_q(\mathfrak{g})\): \(U_q = U \cdot \Phi^{\KZ}_\mathbb{Q}\). We prove this relation in the case of the Lie algebra $\sl_N$ for the tensor cubes of the fundamental and the first symmetric representations. We further check this relation perturbatively for the Lie algebra $\sl_N$ in the representations $[2]\otimes [1] \otimes [1]$, $[2]\otimes [2] \otimes [1]$, the non-trivial cases with multiplicities --- $[2,1]^{\otimes 3}$, $[2,2]^{\otimes 3}$, and in the case of the Lie algebra $\mathfrak{so}_5$ for the tensor cube of the fundamental representation.

\end{abstract}

\tableofcontents

\section{Introduction}

This paper studies the relation between associators and \(6j\) symbols and proposes an explicit formula \(U_q = U\cdot \Phi^{\KZ}_{\mathbb{Q}}\). Let us describe in more detail objects we work with.

The Drinfeld associator $\Phi^{\mathrm{KZ}}(A,B)$
first appeared in Drinfeld's work on quasi-Hopf algebras, where it was constructed as the ratio of two solutions of the Knizhnik-Zamolodchikov (KZ) equation regularized at the
singular points $0$ and $1$~\cite{Drinfeld1990,Drinfeld1991}. Namely,
$\Phi^{\mathrm{KZ}}$ encodes the monodromy of the KZ connection
\[
\frac{dG}{dz} = \hbar\left(\frac{A}{z}+\frac{B}{z-1}\right)G(z),
\]
and satisfies the pentagon and hexagon equations. This places the associator at the heart of the algebraic
structure underlying two-dimensional conformal field theory: it governs the
fusion and braiding of operators in the Wess--Zumino--Witten (WZW)
model~\cite{knizhnikzamolodchikov1984}. Beyond CFT, associators are central to several closely related areas of
mathematical physics and low-dimensional topology. 

Alekseev and Torossian showed
  that every Drinfeld associator produces a solution of the
  Kashiwara--Vergne equations in Lie theory, giving an unexpected bridge
  between associators, the Baker--Campbell--Hausdorff formula, and harmonic
  analysis on Lie groups~\cite{AlekseevTorossian2012,AET2010,ANXZ2018}. 
  
  Furthermore, the coefficients of
  $\Phi^{\mathrm{KZ}}$ are (motivic) multiple zeta values (MZVs)~\cite{lemurakami1996}, and the
  $\mathbb{Q}$-vector space spanned by these coefficients coincides with the
  algebra of MZVs; this identification underlies deep connections between
  associators, periods of mixed Tate motives~\cite{goncharov2001,deligneGoncharov2005,brown2012}, and renormalization in quantum
  field theory, following the pioneering observations of Broadhurst and
  Kreimer on knots and numbers in $\phi^4$-theory~\cite{BroadhurstKreimer1995}.
  
  In string
  perturbation theory, the Drinfeld associator governs the low-energy
  expansion of genus-zero open-string amplitudes, since the
  relevant disk integrals satisfy a KZ-type differential equation in the
  particle number~\cite{Terasoma2014,Kaderli2020}. At genus one, elliptic
  analogs of $\Phi^{\mathrm{KZ}}$ --- elliptic associators built from the
  Knizhnik--Zamolodchikov--Bernard (KZB) connection --- control the
  $\alpha'$-expansion of one-loop amplitudes via elliptic multiple zeta
  values~\cite{BroedelMatthesSchlotterer2015}. In parallel, modern
  approaches to Feynman integrals based on canonical differential equations
  and the theory of iterated integrals use the Drinfeld associator to
  transport boundary data between kinematic regimes, reproducing the
  well-known appearance of multiple zeta values in the epsilon-expansion of
  dimensionally-regularized integrals~\cite{HennLectures2015}.
  
The relation between associators and knot theory is of interest as well. Let us consider knot invariants arising as averages of Wilson loops in the topological Chern--Simons theory~\cite{CS,Witten} with gauge group \(G\):
\begin{equation}
     \label{WilsonLoopExpValue}
\begin{aligned}
             &S_{\text{CS}}[{\cal A}] = \frac{\kappa}{4 \pi} \int_{S^3} \text{tr} \left( {\cal A} \wedge d{\cal A} +  \frac{2}{3} {\cal A} \wedge {\cal A} \wedge {\cal A} \right)\,,\\
             &\langle W(K)\rangle = \left\langle \text{tr}_{R} \ P \exp \left( \oint_{{K}} {\cal A} \right) \right\rangle_{\text{CS}}\,. 
\end{aligned}
\end{equation}
To compute these averages, we need to fix some gauge. In the holomorphic gauge case, the vacuum expectation value is equal to the Kontsevich integral \cite{chmutovduzhinmostovoy2011,kontsevich1993,labastidaperez1997}. Alternatively, the Kontsevich integral can be computed in another way, by taking, in some representation \(R\) of the group \(G\), the trace of the product of the \(\mathcal{R}\)-matrix operators and the associators \(\Phi\)~\cite{piunikhin1993,lemurakami1994,chmutovduzhinmostovoy2011}.\\

Recoupling coefficients, or $6j$-symbols, originate in the quantum theory of
angular momentum, where they express the change of basis between two
inequivalent ways of decomposing a triple tensor product of $\mathfrak{su}(2)$
representations~\cite{Racah1942}. Algebraically, they encode the associativity
constraint for the fusion of representations, so that $6j$-symbols, in
essence, play a role similar to that of an associator. Their quantum deformation for
${\cal U}_q(\mathfrak{g})$ satisfies its own pentagon and hexagon
identities. They further arise as
fusion matrices of WZW conformal blocks~\cite{moore1989classical} and, more
recently, in holographic computations of conformal blocks and Virasoro
$6j$-symbols in AdS/CFT~\cite{TeschnerVartanov2014,LiuEtAl2019}.

While the $\mathfrak{su}(2)$ case is completely understood -- an explicit
${}_4F_3$ hypergeometric expression was found already by
Racah~\cite{Racah1942}, together with its full symmetry group (tetrahedral
symmetries and the Regge symmetry) --- the structure of $6j$-symbols for
higher-rank algebras $\mathfrak{sl}_N$, $N>2$, and for other classical series
remains an active research subject~\cite{hecht1975simple, gustafson1987whipple, wong1979multiplicity, artamonov2025calculation}. Explicit multiple-sum formulas are known
only for special (e.g. symmetric or rectangular) representations, and even
basic symmetries beyond the tetrahedral ones were unknown until
recently~\cite{alisauskas2002}. A recent line of work by Morozov, Sleptsov and
collaborators, based on the \emph{eigenvalue hypothesis}~\cite{itoyama2013eigenvalue} --- the conjecture
that ${\cal U}_q(\mathfrak{sl}_N)$ Racah matrices are uniquely fixed by the
eigenvalues of the associated quantum $\mathcal{R}$-matrices --- has produced
new families of symmetries (the ``tug-the-hook'' symmetry and others) for
${\cal U}_q(\mathfrak{sl}_N)$ $6j$-symbols with symmetric representations, and proved
the hypothesis itself in the ${\cal U}_q(\mathfrak{sl}_2)$
case~\cite{MorozovSleptsov2018,AlekseevMorozovSleptsov2019,LaninaSleptsov2022}.

In addition, they play a major role in computing quantum knot invariants. Let us return to the averages of Wilson loops in Chern-Simons theory (\ref{WilsonLoopExpValue}). If one chooses the temporal gauge, then these averages will be proportional to quantum knot polynomials~\cite{smirnov2009,morozovsmirnov2010}, such as, for example, the HOMFLY--PT~\cite{freyd1985new,przytycki1988invariants} polynomial or the Kauffman~\cite{kauffman1990invariant} polynomial. To compute these polynomials, there exists the Reshetikhin--Turaev technique  \cite{reshetikhinturaev1990,reshetikhinturaev1991}, which can be reduced to computing traces of the product of two operators: the \(\mathcal{R}\)-matrix (which differs from the \({\cal R}\)-matrix in the computation of the Kontsevich integral) and the quantum \(6j\)-symbols.


\bigskip

The paper is organized as follows. In Section~\ref{sec::2}, we give the basic definitions from the theory of quasi-Hopf algebras and also explain the notion of \(6j\)-symbols for an arbitrary quasi-Hopf Lie algebra and why they should be related to the associator. Section~\ref{sec::3} is devoted to verifying our conjectural formula, and we briefly clarify what the object \(\Phi^{KZ}_{\mathbb{Q}}\) is. In Section~\ref{sec::4}, we briefly interpret our results and propose possible directions for future research. In Appendix~\ref{sec:A}, we analyze an example of a twist transforming a quasi-Hopf algebra into the deformation quantization of \(\mathfrak{sl}_2\). Appendix~\ref{sec:B} shows how to compute associators from the Knizhnik--Zamolodchikov equation for the tensor cube of the fundamental and symmetric representations of the algebra \(\mathfrak{sl}_2\). In Appendix~\ref{sec:C}, we list the matrices we have used to verify our formula.

\section{Quasi-Hopf algebras}
\label{sec::2}
A Hopf algebra is a pair consisting of an associative algebra $A$ and a comultiplication on it $\Delta: A \to A \otimes A$. The presence of this comultiplication allows us to introduce the tensor product of representations of our algebra, $V_1 \otimes V_2$. It is then natural to require ``good'' properties for the product of representations --- namely, coassociativity and cocommutativity --- so that we have the following natural isomorphisms:
\begin{equation}\label{repassoc}
    (V_1 \otimes V_2) \otimes V_3 \cong V_1 \otimes (V_2 \otimes V_3)\,, 
\end{equation}
\begin{equation}\label{repcomm}
    V_1 \otimes V_2 \cong V_2 \otimes V_1\,,
\end{equation}
which imposes the following constraints on the homomorphism $\Delta$:
\begin{equation}
    (\Delta \otimes \mathrm{id}) \circ \Delta = (\mathrm{id} \otimes \Delta) \circ \Delta\,,
\end{equation}
\begin{equation}
    \Delta' = \sigma \circ \Delta = \Delta, \quad \sigma(a \otimes b) = b \otimes a\,.
\end{equation}
However, as is well known, the cocommutativity condition can be relaxed by introducing an $\CR$-matrix, which leads to the notion of a quasitriangular Hopf algebra $(A, \Delta, {\cal R})$, where ${\cal R} \in A \otimes A$:
\begin{equation}
    \Delta'(a) = {\cal R} \Delta(a) {\cal R}^{-1}\,, 
\end{equation}
then, the isomorphism (\ref{repcomm}) still exists, but its definition involves the $\CR$-matrix. Of course, one must impose a self-consistency condition on the element $\CR$, which leads to the Yang-Baxter equation:
\begin{equation}
    {\cal R}_{12}{\cal R}_{13}{\cal R}_{23}={\cal R}_{23}{\cal R}_{13}{\cal R}_{12}\,.
\end{equation}
It is natural to try to weaken the coassociativity condition as well, which leads to the notion of an associator $\Phi \in A \otimes A \otimes A$ and a quasi-Hopf algebra:
\begin{equation}
    (\mathrm{id} \otimes \Delta) \Delta(a) = \Phi \, (\Delta \otimes \mathrm{id}) \Delta(a) \, \Phi^{-1}, \quad a \in A\,.
\end{equation}
Such an algebra is still suitable for constructing knot invariants.
Moreover, such an algebra retains gauge freedom, described by transformations called twists $\mathcal{F} \in A \otimes A$.\\[6pt]
In this section, we recall the main definitions and properties of quasi-Hopf algebras, illustrating them with examples whenever possible. In particular, we shall focus on the Drinfeld quantization of Lie algebras.

\subsection{Quasi-Hopf Lie algebra}

A quasi-Hopf algebra is a Hopf algebra in which the coassociativity axiom is replaced by a weaker condition. 
\begin{Def}
    A quasi-Hopf algebra over a ring $\mathbb{K}$ is a set $(A,\Delta,\varepsilon,\Phi)$, where $A$ is an associative $\mathbb{K}$-algebra with unit, $\Delta: A \rightarrow A \otimes A$ is a homomorphism (the comultiplication), $\varepsilon: A \rightarrow \mathbb{K}$ is a homomorphism (counit), $\Phi \in A \otimes A \otimes A$ is an invertible element (the associator), and $S: A \rightarrow A$ is an antihomomorphism (the antipode), such that:
    \begin{equation}\label{ax1}
        (\id\otimes\Delta)\Delta(a)=\Phi(\Delta\otimes \id)\Delta(a)\Phi^{-1},\quad a\in A\,,
    \end{equation}
    \begin{equation}\label{ax2}
        (\id\otimes \id \otimes \Delta)(\Phi)(\Delta\otimes \id \otimes \id) (\Phi) = (\mathds{1}\otimes\Phi)(\id\otimes\Delta\otimes \id)(\Phi)(\Phi\otimes \mathds{1})\,,
    \end{equation}
    \begin{equation}
        (\varepsilon\otimes \id)\Delta=(\id \otimes \varepsilon)\Delta = \id\,,
    \end{equation}
    \begin{equation}
        (\id \otimes \varepsilon \otimes \id)(\Phi)=\mathds{1}\,,
    \end{equation}
\end{Def}

Now, in contrast to ordinary Hopf algebras, the associativity isomorphism for representations is not the identity. If \(R_1, R_2, R_3\) are some representations of given algebra, then:
\begin{equation}\label{iso}
    (R_1\otimes R_2) \otimes R_3 \overset{\Phi}{\cong} R_1 \otimes (R_2 \otimes R_3)\,.
\end{equation}
One can further define a weakened notion of a quasitriangular Hopf algebra:
\begin{Def}
    A quasitriangular quasi-Hopf algebra is a set $(A,\Delta,\varepsilon,\Phi, R)$, where $(A,\Delta,\varepsilon,\Phi)$ is a quasi-Hopf algebra and $\CR \in A \otimes A$ is an invertible element such that:
    \begin{equation}\label{ax4}
        \Delta'(a) = \sigma\Delta(a)= {\cal R}\Delta(a){\cal R}^{-1}\,, 
    \end{equation}
    \begin{equation}\label{ax5}
        (\Delta\otimes \id)({\cal R})=\Phi^{321}{\cal R}^{13}(\Phi^{132})^{-1}{\cal R}^{23}\Phi\,,
    \end{equation}
    \begin{equation}\label{ax6}
        (\id\otimes\Delta)({\cal R})=(\Phi^{231})^{-1}{\cal R}^{13}\Phi^{213}{\cal R}^{12}\Phi^{-1}\,,
    \end{equation}
    where $\sigma: A \otimes A \rightarrow A \otimes A$ is the flip of tensor factors, and the following notation is introduced: if $X = a \otimes b$, then, for example, $X^{12} = a \otimes b \otimes 1$ and $X^{13} = a \otimes 1 \otimes b$; and if $X = a \otimes b \otimes c$, then $X^{312} = c \otimes a \otimes b$.
\end{Def}
Let us note that for a quasi-Hopf quasitriangular algebra there is still a Yang-Baxter equation, but now involving the associator:
\begin{equation}
    {\cal R}_{12}\Phi_{312}{\cal R}_{13}\Phi^{-1}_{132}{\cal R}_{23}\Phi_{123}=\Phi_{321}{\cal R}_{23}\Phi^{-1}_{231}{\cal R}_{13}\Phi_{213}{\cal R}_{12}
\end{equation}
Let us now define quasi-Hopf Lie algebras:
\begin{Def}
    Let \(\mathbb{K}\) be a field of characteristic \(0\). A quasitriangular quasi-Hopf QUE (quantized universal enveloping) algebra over \(\mathbb{K}[[h]]\) is a topological quasitriangular quasi-Hopf algebra \((A, \Delta, \varepsilon, \Phi, {\cal R})\) over \(\mathbb{K}[[h]]\) such that \(A/hA\) is a universal enveloping algebra with the standard comultiplication, and \(A\), as a topological \(\mathbb{K}[[h]]\)-module, is isomorphic to \(V[[h]]\) for some vector space \(V\) over \(\mathbb{K}\).
\end{Def}
Quasi-Hopf algebras admit gauge transformations, called twists, which send one algebra to another but with equivalent categories of representations as quasitensor categories.
\begin{Def}
    A twist $\mathcal{F} \in A \otimes A$ is an invertible element. The twisted algebra $A^{\mathcal{F}} = (A, \Delta^{\mathcal{F}}, \varepsilon, \Phi^{\mathcal{F}},  {\cal R}^{\mathcal{F}})$ is defined as follows:
    \begin{equation}
        \Delta^{\mathcal{F}}(a) = \mathcal{F} \Delta(a) \mathcal{F}^{-1}
    \end{equation}
    \begin{equation}
        \Phi^{\mathcal{F}} = \mathcal{F}^{12} (\Delta\otimes id)(\mathcal{F})\Phi (id\otimes \Delta)(\mathcal{F})^{-1} (\mathcal{F}^{23})^{-1}
    \end{equation}
    \begin{equation}
         {\cal R}^{\mathcal{F}} = \mathcal{F}^{21}\cdot  {\cal R}\cdot (\mathcal{F}^{12})^{-1}
    \end{equation}
\end{Def}
There is the following result due to V. Drinfeld describing all quasi-Hopf Lie algebras.
\begin{Th}
    Every quasitriangular quasi-Hopf Lie algebra can be transformed, via a twist, into the quantum group $\mathcal{U}_{q}(\mathfrak{g})$.
\end{Th}
Constructing such a twist explicitly, however, is notoriously
difficult. At present, no closed-form expression for the universal twist  is known for any rank higher than the
simplest cases~\cite{blohmann2005reconstruction, kulish2006twists}. An example of constructing the first-order of such a twist
transforming \(\mathcal{U}(\mathfrak{sl}_2)[[h]]\) into \(\mathcal{U}_q(\mathfrak{sl}_2)\) can be found in Appendix~\ref{sec:A}.

\subsection{6j-symbols}
Let us introduce the \(6j\)-symbols for an arbitrary quasi-Hopf Lie algebra. Suppose we have the tensor product of three representations \(R_1 \otimes R_2 \otimes R_3\); as is well known, it can be decomposed into irreducible representations in two different ways:
\begin{align}
    (R_1 \otimes R_2) \otimes R_3 = \bigoplus_{j} R_j \otimes R_3 = \bigoplus_{j,k} R_{jk}\,, \\
    R_1 \otimes (R_2 \otimes R_3) = \bigoplus_{i} R_1 \otimes R_i = \bigoplus_{i,l} R_{li}\,.
\end{align}
The decomposition into irreducible representations is determined by the operators \((\id \otimes \Delta)\Delta(a)\) and \((\Delta \otimes \id)\Delta(a)\) and depends directly on which comultiplication is chosen in the algebra.

Now let us define the \(6j\)-symbols, not forgetting about the associativity isomorphism (\ref{iso}), by means of the following commutative diagram:
\begin{equation}
    \begin{tikzcd}
        (R_1 \otimes R_2) \otimes R_3 \arrow{rr}{C_L} \arrow{dd}{\Phi} &  & \bigoplus_{\lambda} R_{\lambda} \arrow{dd}{\tilde{U}} \\
                                          &  &                   \\
        R_1 \otimes (R_2 \otimes R_3) \arrow{rr}{C_R}                 &  & \bigoplus_{\mu } R_{\mu}                
    \end{tikzcd}
\end{equation}
where \(C_L\) and \(C_R\) are the basis-change matrices corresponding to different decompositions into irreducible representations. Then the \(6j\) symbols are defined as:
\begin{equation}
    \tilde{U} = C_R \cdot \Phi \cdot C_L^{-1}\,.
\end{equation}
Let us note that if we have some Lie algebra \(\mathfrak{g}\) with a trivial associator $\Phi = \mathds{1}$, then one can obtain the definitions of two standard \(6j\)-symbols:
\begin{itemize}
\item For the classical comultiplication \(\Delta(a) = \mathds{1} \otimes a + a \otimes \mathds{1}\), we have the classical \(6j\)-symbols of a Lie algebra \(\mathfrak{g}\): \(U = C_R C_L^{-1}\).
\item For the deformed comultiplication \(\Delta_q(a)\), we have the quantum \(6j\)-symbols of the algebra \(\mathcal{U}_q(\mathfrak{g})\): \(U_q = C_R^q (C_L^{q})^{-1}\).
\end{itemize}
Now, let us write the associator \(\Phi\) in the basis corresponding to the left decomposition into irreducible representations, that is, \(\Phi \to C_L^{-1} \Phi C_L\). Then:
\begin{equation}
    \tilde{U} = C_R C_L^{-1}\cdot \Phi\,.
\end{equation}
Let us note that if we have the standard comultiplication \(\Delta(a) = \mathds{1} \otimes a + a \otimes \mathds{1}\) and the standard generators of the Lie algebra \(\mathfrak{g}\) (as, for example, in the quasi-Hopf algebra \(\mathcal{U}(\mathfrak{g})[[h]]\)), then one can use the classical \(6j\)-symbols from the matrices \(C_L\) and \(C_R\):
\begin{equation}
    U = C_RC_L^{-1}\,,
\end{equation}
\begin{equation}
    \tilde{U} = U \cdot \Phi\,.
\end{equation}
From Drinfeld's theorem, as we know, all such quasi-Hopf Lie algebras are equivalent to the algebra \(\mathcal{U}_q(\mathfrak{g})\), and hence their \(6j\) symbols must likewise be related by a twist:
\begin{equation}\label{tw}
    U_q \overset{\mathcal{F}}{\longleftrightarrow} \tilde{U} = U \cdot \Phi\,.
\end{equation}
However we have not found this relation explicitly anywhere. We will also not describe this relation explicitly, but instead propose the following empirical formula:
\begin{equation}
U_q = U\cdot \Phi^{\KZ}_{\mathbb{Q}}\,.
\end{equation}

\subsection{Quantization of Lie algebras}
Following Drinfeld \cite{drinfeld1989quasi}, let us describe a method for constructing quasi-Hopf quasitriangular QUE algebras.\\
Let us define \(A = \mathcal{U}(\mathfrak{g})[[h]]\) (as an algebra over formal power series in \(h\)) and the standard comultiplication on it, \(\Delta(a) = \mathds{1} \otimes a + a \otimes \mathds{1}\). Suppose there is a symmetric \(\mathfrak{g}\)-invariant tensor \(\tau \in \mathfrak{g} \otimes \mathfrak{g}\) (the split Casimir operator), and set \(\mathcal{R} = \exp\left(\frac{1}{2} h \tau\right)\). For the associator, there is the following theorem, also due to Drinfeld:
\begin{Th}
    For a given algebra $\mathfrak{g}$ with such a comultiplication and $\CR$-matrix, there exists a unique associator up to twist by symmetric $\mathfrak{g}$-invariant $\mathcal{F}$.
\end{Th}
Furthermore, there is a theorem on the existence of a rational associator:
\begin{Th}
There exists a \(\mathbb{Q}\)-universal formula expressing \(\Phi\) in terms of \(\tau\). It is unique up to symmetric twists.
\end{Th}
However, for such rational associators no explicit formula exists. We, in turn, will describe a complete construction for the transcendental associator \(\Phi^{\KZ}\) by means of the Knizhnik-Zamolodchikov equations (henceforth we will always have \(\mathbb{K} = \mathbb{C}\)).\\
 Let us consider the four-point correlation function in the WZW model for a Lie algebra $\mathfrak{g}$ with three fixed points:
\begin{equation}
    G(z) = \langle\varphi(0)\varphi(1)\varphi(\infty)\varphi(z)\rangle\,.
\end{equation}
Then, one can write the Knizhnik--Zamolodchikov equation for it:
\begin{equation}\label{kz}
    \frac{dG(z)}{dz} = \hbar \left( \frac{A}{z} - \frac{B}{1-z} \right) G(z)\,,
\end{equation}
where the matrices $A$ and $B$ are the split Casimir operators of $\mathfrak{g}$:
\begin{equation}
    A = t_a \otimes t_a^* \otimes \mathbf{1}, \qquad B = \mathbf{1}\otimes t_a \otimes t_a^*\,.
\end{equation}
Note that here:
\begin{equation}
    \hbar = \frac{h}{2\pi i}\,.
\end{equation}
The Drinfeld associator is defined as:
\begin{equation}
    \Phi = G_1^{-1}(z) G_0(z)\,.
\end{equation}
Here $G_i(z)$ are the following asymptotics:
\begin{equation}
    G_0(z) \sim z^{\hbar A},\; z\rightarrow 0, \qquad G_1(z) \sim (1-z)^{\hbar B},\; z\rightarrow 1\,.
    \label{eq:asymp-rel}
\end{equation}
Let us note that \(\Phi\) does not depend on \(z\), since \(G_0\) and \(G_1\) are two solutions of the same equation on the interval \(0 < z < 1\).

We can give an equivalent definition of the associator that is more convenient for computation. Let us remove the divergences at \(z = 0, 1\) by means of the substitution
\begin{equation}
\tilde{\Phi}(z) = (1-z)^{-\hbar B} G(z) z^{-\hbar A}, \quad \tilde{\Phi}(0) = 1\,.
\end{equation}
Then we obtain the following regularized equation:
\begin{equation}
    \frac{d\tilde{\Phi}}{dz} = \frac{\hbar}{z}(1-z)^{-\hbar B} A (1-z)^{\hbar B}\tilde{\Phi} - \frac{\hbar}{z}\tilde{\Phi} A,\qquad\text{with } \tilde{\Phi}(0)=1\,.
    \label{eq:eq-on-Phi}
\end{equation}
It is not hard to verify that the value of the solution of this equation at \(z = 1\) is the associator:
\begin{equation}
    \Phi = \tilde{\Phi}(z=1)\,.
\end{equation}
The set $(\mathcal{U}(\mathfrak{g})[[\hbar]], \Delta, \Phi, \CR)$ forms a quasitriangular quasi-Hopf algebra. Axioms (\ref{ax1}) and (\ref{ax4}) are clearly satisfied because the associator and the $\CR$-matrix come from Casimir operators, while (\ref{ax2}) is the pentagon relation for $\Phi$.

For the Drinfeld associator, there is a formula in the form of a series over all possible commutators and zeta values, obtained by Le and Murakami in \cite{lemur}.

\begin{equation}\label{Drseries}
    \Phi_{\mathrm{KZ}} = 1 + \sum_{m=2}^{\infty} \hbar^m
    \sum_{\substack{0 < \mathbf{p},\ 0 < \mathbf{q} \\ |\mathbf{p}| + |\mathbf{q}| = m}} 
    \eta(\mathbf{p}, \mathbf{q}) \cdot 
    \sum_{\substack{0 \leq \mathbf{r} \leq \mathbf{p} \\ 0 \leq \mathbf{s} \leq \mathbf{q}}} 
    (-1)^{|\mathbf{r}| + |\mathbf{s}|} 
    \binom{\mathbf{p}}{\mathbf{r}} \binom{\mathbf{q}}{\mathbf{s}} \cdot 
    B^{|\mathbf{s}|} (A, B)^{(\mathbf{i}, \mathbf{j})} A^{|\mathbf{r}|}.
\end{equation}

The following notation is introduced here:
\[
\eta(\mathbf{p}, \mathbf{q}) := 
\zeta\!\left(
\underbrace{1, \dots, 1}_{q_l-1},\,
p_l + 1,\,
\underbrace{1, \dots, 1}_{q_{l-1}-1},\,
p_{l-1} + 1,\,
\dots,\,
\underbrace{1, \dots, 1}_{q_1-1},\,
p_1 + 1
\right).
\]

Furthermore, we set
\[
\mathbf{r} = (r_1, \ldots, r_l); \quad 
\mathbf{i} = (i_1, \ldots, i_l); \quad 
\mathbf{s} = (s_1, \ldots, s_l); \quad 
\mathbf{j} = (j_1, \ldots, j_l);
\]
\[
\mathbf{p} = \mathbf{r} + \mathbf{i} = (r_1 + i_1, \ldots, r_l + i_l); \quad 
\mathbf{q} = \mathbf{s} + \mathbf{j} = (s_1 + j_1, \ldots, s_l + j_l);
\]
\[
|\mathbf{r}| = r_1 + \cdots + r_l; \quad 
|\mathbf{s}| = s_1 + \cdots + s_l;
\]
\[
\binom{\mathbf{P}}{\mathbf{R}}
=
\binom{p_1}{r_1}
\binom{p_2}{r_2}
\cdots
\binom{p_l}{r_l};
\quad
\binom{\mathbf{Q}}{\mathbf{S}}
=
\binom{q_1}{s_1}
\binom{q_2}{s_2}
\cdots
\binom{q_l}{s_l};
\]
\[
(A, B)^{(\mathbf{i}, \mathbf{j})} = A^{i_1} B^{j_1} \cdots A^{i_l} B^{j_l}.
\]
For further convenience, we write out the first few orders of this series.
\begin{equation}
\begin{aligned}
    \Phi &= \mathds{1} -\hbar^2 \zeta(2) [A,B] - \zeta(3)\hbar^3 [A+B,[A,B]] + \dots
\end{aligned}
\end{equation}

\section{Associators vs 6j-symbols}
\label{sec::3}
Associators and $6j$-symbols play closely related roles, suggesting that there should be a natural relation between these two objects. We propose the following conjectural formula relating them:
\begin{equation}\label{hyp}
    \boxed{U_q = U \cdot \Phi^{\KZ}_{\mathbb{Q}}}.
\end{equation}
It should be emphasized that the associator $\Phi^{\KZ}_{\mathbb{Q}}$ is expressed in the basis corresponding to the bracketing \((R_1 \otimes R_2)\otimes R_3,\)
whereas the $6j$-symbols describe the recoupling transformation between the two natural bases associated with the two possible bracketings of the triple tensor product,
\[
(R_1 \otimes R_2) \otimes R_3
\qquad\text{and}\qquad
R_1 \otimes (R_2 \otimes R_3).
\]
The new object $\Phi^{\KZ}_{\mathbb{Q}}$ appearing in \eqref{hyp} will be referred to as the ``rationalized Drinfeld associator''. We now explain the meaning of this terminology by formulating two hypotheses.\\
Our first hypothesis is motivated by the explicit computations presented in the following sections.
\begin{conjecture}
Let $\Phi^{\KZ}$ denote the Drinfeld associator obtained from the Knizhnik--Zamolodchikov equation, or equivalently from the expansion \eqref{Drseries}. Its coefficients are, in general, expressed in terms of multiple zeta values. We define the rationalized Drinfeld associator $\Phi^{\KZ}_{\mathbb{Q}}$ by retaining, in the expansion of $\Phi^{\KZ}$, only those terms whose coefficients are proportional to a single even zeta value. We then conjecture that \(U_q = U \cdot \Phi^{\KZ}_{\mathbb{Q}}\).

\end{conjecture}
We can formulate a more general conjecture which, for the examples considered in this paper, is equivalent to Conjecture 1.
\begin{conjecture}
Let $\Phi^{\KZ}$ be the Drinfeld associator obtained from the Knizhnik--Zamolodchikov equation or, equivalently, from the expansion \eqref{Drseries}. Its coefficients are expressed in terms of multiple zeta values. We define $\Phi^{\KZ}_{\mathbb{Q}}$ to be the associator obtained from $\Phi^{\KZ}$ by eliminating all contributions involving irrational zeta values, using the algebraic relations among multiple zeta values. Then \(U_q = U \cdot \Phi^{\KZ}_{\mathbb{Q}}\).
\end{conjecture}
A more detailed description of this rationalized associator will be given in the following section.\\
We further note that, up to order $\hbar^7$, both of the rationalized associators defined above coincide with the Bar-Natan associator $\Phi^{\mathrm{BN}}$, which is constructed in \cite{barnatan} by successively solving the pentagon and hexagon equations \eqref{ax2}, \eqref{ax5}, and \eqref{ax6}.\\
In this section, we shall verify the hypothetical connection between the associator and the $6j$-symbols.

\subsection{Rational associator}

In the present work, the term ``rationality'' is used in a specific sense when referring to zeta values and their appearance in the Drinfeld associator. We therefore find it appropriate to clarify this notion.\\
We begin by noting that, in the Drinfeld associator arising from the Knizhnik--Zamolodchikov equation, after the substitution $\hbar = \frac{h}{2\pi i}$ in formula (\ref{Drseries}), the zeta values enter through expressions of the form
\begin{equation}
\frac{\zeta(k_1, \dots, k_r)}{\pi^{n}} , \qquad n = \sum_{i=1}^r k_i,
\end{equation}
where $n$ is the weight of the multiple zeta value.

Consequently, in order for the coefficients of the associator to be rational numbers, we adopt the following convention: a multiple zeta value $\zeta(k_1, \dots, k_r)$ of weight $n$ will be called \emph{rational} if:
\begin{equation}
\zeta(k_1, \dots, k_r) \in \pi^n\mathbb{Q}.
\end{equation}
In this context, even zeta values are rational:
\[
\zeta(2n) \in \pi^{2n} \mathbb{Q}.
\]
The odd zeta values, by contrast, are much more difficult to understand. Apéry proved~\cite{apery1979} that $\zeta(3)$ is irrational, but its transcendence is unknown. For $\zeta(5), \zeta(7), \dots$, not even irrationality is known in general; however, Zudilin \cite{zudilin2001} showed that at least one of $\zeta(5), \zeta(7), ...$ is irrational, and Rivoal \cite{rivoal2000} established that infinitely many of the odd zeta values $\zeta(2m+1)$ $(m \ge 2)$ are irrational, without identifying which.

Nonetheless, for the purposes of the present work, we adopt the following standard conjecture:

\begin{conjecture}[Transcendence conjecture]
The set 
\[
\{\pi,\ \zeta(3),\ \zeta(5),\ \zeta(7),\ \dots\}
\]
is algebraically independent over $\mathbb{Q}$.
\end{conjecture}

Under this assumption, the odd zeta values are treated as formal transcendental indeterminates, and any term involving them must be discarded in the rationalization procedure.

A multiple zeta value of weight $n$ and depth $r$ is defined as
\[
\zeta(k_1, \dots, k_r) = \sum_{n_1 > \cdots > n_r > 0} \frac{1}{n_1^{k_1} \cdots n_r^{k_r}},
\]
where $k_i \in \mathbb{Z}_{>0}$, $k_1 \ge 2$, and $n = k_1 + \cdots + k_r$. 
At present, not all relations among multiple zeta values are known; however, it is conjectured that all of them are generated by the double shuffle relations, which are described in detail in the literature (see for example \cite{ikz2006},\cite{racinet2002}).

As we can see, among the basis elements there appear, for example, multiple zeta values such as $\zeta(6,2)$, which we assume to be independent of the others. Consequently, we discard them in the rationalization procedure.\\
We assume that the object \(\Phi^{\KZ}_{\mathbb{Q}}\) which we have constructed is an associator. One can split the Drinfeld associator into a rational and a transcendental part,
\[
\Phi^{\KZ} = \Phi^{\KZ}_{\mathbb{Q}} + \Phi^{\KZ}_{T},
\]
and substitute it into the pentagon and hexagon relations  (\ref{ax2}),(\ref{ax5}),(\ref{ax6}), and then split the equations into a rational and a transcendental part, each of which must vanish separately and yield a new pentagon and hexagon relation, now for \(\Phi^{\KZ}_{\mathbb{Q}}\).\\

Moreover, we do so because it resembles another rational associator introduced by Bar-Natan in his paper \cite{barnatan}. He gave a combinatorial description of the construction of the pair \((\CR, \Phi^{\rm BN})\) by successively solving the pentagon and hexagon equations for the \(\CR\)-matrix and the associator, and it turns out that all the coefficients of this pair are rational. One can compare the associators \(\Phi^{\KZ}_{\mathbb{Q}}\) and \(\Phi^{\rm BN}\) and observe that, at least up to order 7 (the order to which the computation was carried out in \cite{barnatan}), they coincide completely.

\subsection{Exact verification for $[1]^{\otimes 3}$ of the $\mathfrak{sl}_N$ algebra}
\label{subsec:exac-fund-sl_N}

Let us prove that this formula holds exactly for the tensor cube of the fundamental representation $\Box^{\otimes 3}$ of $\mathfrak{sl}_N$.

We shall need to properly rationalize the Drinfeld associator obtained from the Knizhnik--Zamolodchikov equation (see the derivation in Appendix~\ref{sec:B}):
\begin{equation}
    \Phi= \frac{1}{4}\left(\begin{array}{ c  c }
    \displaystyle\frac{\Gamma(1+2\hbar)\Gamma(1-2\hbar)}{\Gamma(1+\hbar)\Gamma(1-\hbar)}+3\displaystyle\frac{\Gamma^2(1+2\hbar)}{\Gamma(1+3\hbar)\Gamma(1+\hbar)}   & -\sqrt{3}\left(\displaystyle\frac{\Gamma^2(1-2\hbar)}{\Gamma(1-3\hbar)\Gamma(1-\hbar)}-\frac{\Gamma(1-2\hbar)\Gamma(1+2\hbar)}{\Gamma(1+\hbar)\Gamma(1-\hbar)}\right) \\
     -\sqrt{3}\left(\displaystyle\frac{\Gamma(1+2\hbar)\Gamma(1-2\hbar)}{\Gamma(1+\hbar)\Gamma(1-\hbar)}-\displaystyle\frac{\Gamma^2(1+2\hbar)}{\Gamma(1+3\hbar)\Gamma(1+\hbar)}\right) & 3\displaystyle\frac{\Gamma^2(1-2\hbar)}{\Gamma(1-3\hbar)\Gamma(1-\hbar)}+\frac{\Gamma(1-2\hbar)\Gamma(1+2\hbar)}{\Gamma(1+\hbar)\Gamma(1-\hbar)}
  \end{array}\right).
\end{equation}
We use the Taylor expansion of the gamma function and separate the sum into even and odd zeta values:
\begin{equation}\label{gamma}
     \Gamma(1+z)=\exp\left(-\gamma z + \sum_{n=2}^{\infty} \frac{(-1)^{n}}{n}\zeta(n)z^n\right)=\exp\left(-\gamma z + \sum_{n=1}^{\infty}\frac{\zeta(2n)}{2n}z^{2n} - \sum_{k=1}^{\infty}\frac{\zeta(2k+1)}{2k+1}z^{2k+1}\right)\,.
\end{equation}
Recall that the even zeta values are given by
\begin{equation}
\zeta(2n) = \frac{(-1)^{n+1}  B_{2n}}{2(2n)!}(2\pi)^{2n},
\end{equation}
where $B_{2n}$ are the Bernoulli numbers. Upon substituting $z = \hbar = \frac{h}{2 i\pi}$ into formula (\ref{gamma}), all powers of $\pi$ accompanying the even zeta values cancel out.

As for the odd zeta values, we assume that $\{\pi, \zeta(3), \zeta(5), \dots\}$ are algebraically independent over $\mathbb{Q}$; consequently, we discard them from our associator.

From the Euler product formula for the sine function,
\begin{equation}
\frac{\sin \pi z}{\pi z} = \prod_{n=1}^{\infty} \left(1 - \frac{z^2}{n^2}\right),
\end{equation}
we obtain the following expansion of the sine in terms of even zeta values:
\begin{equation}
\frac{\sin \pi z}{\pi z} = \exp\!\left( - \sum_{n=1}^{\infty} \frac{\zeta(2n)}{n} z^{2n} \right).
\end{equation}
from which we obtain:
\begin{equation}
    \log \Gamma(1+z)=-\gamma z-\frac{1}{2}\log \frac{\sin(\pi z)}{\pi z}-\sum_{k=1}^{\infty}\frac{\zeta(2k+1)}{2k+1}z^{2k+1}\,.
\end{equation}
Having substituted these expansions into the associator, we then apply the rationalization procedure by removing all contributions involving odd zeta values, and obtain:
\begin{equation}
    \Phi^{\KZ}_{\mathbb{Q}} =\frac{1}{2} \begin{pmatrix}
        \frac{1}{[2]_q} + \sqrt{3}\frac{\sqrt{[3]_q}}{[2]_q} & \sqrt{3}\frac{1}{[2]_q} + \frac{\sqrt{[3]_q}}{[2]_q}\\
        -\sqrt{3}\frac{1}{[2]_q} + \frac{\sqrt{[3]_q}}{[2]_q} & \frac{1}{[2]_q} - \sqrt{3}\frac{\sqrt{[3]_q}}{[2]_q}
    \end{pmatrix},
\end{equation}
where $q=e^{2h}$ and $[n]_q = \frac{q^{n}-q^{-n}}{q - q^{-1}}$.

Thus, we see that the formula $U_q = U\cdot \Phi^{\mathrm{KZ}}_{\mathbb{Q}}$ holds to all orders in $h$.

\subsection{Exact verification for $[2]^{\otimes 3}$ of the $\mathfrak{sl}_N$ algebra}
\label{subsec:exac-sym-sl_N}

One can solve the KZ equation in the case of $\sl_N$ Lie algebra for the tensor cube of the first symmetric representation, see Appendix~\ref{sec:B}. The answer for the associator is: 

\begin{multline}\label{UPhi-[2]^3}
    U \cdot \Phi =\\
    =\begin{pmatrix}
        \displaystyle\frac{\sec(2\pi\hbar)}{2+4\cos(2\pi\hbar)} & \displaystyle\frac{2^{-6\hbar}\Gamma(1-3\hbar)\Gamma^2(\frac{1}{2}-2\hbar)\Gamma(2\hbar)}{\sqrt{15\pi}\Gamma(\frac{1}{2}-\hbar)\Gamma(-5\hbar)\Gamma(1+\hbar)} & \displaystyle\frac{\sqrt{5}\Gamma(1-3\hbar)\Gamma(-3\hbar)}{\Gamma(1-5\hbar)\Gamma(-\hbar)}\\
        \displaystyle\frac{2^{4\hbar}\sqrt{15}\hbar\Gamma(\frac{1}{2}-\hbar)\Gamma(3\hbar)\Gamma^2(\frac{1}{2}+2\hbar)}{2\pi\Gamma(\frac{1}{2}+\hbar)\Gamma(1+5\hbar)} & \displaystyle\frac{\sec(2\pi\hbar)-2}{2} & \displaystyle-\frac{2^{-2\hbar}\Gamma(1-3\hbar)\Gamma^2(\frac{1}{2}-\hbar)\Gamma(\frac{1}{2}+2\hbar)}{\sqrt{3}\pi\Gamma(1-4\hbar)\Gamma(\frac{1}{2}+\hbar)}\\
        \displaystyle\frac{\sqrt{5}\Gamma^2(1+3\hbar)}{3\Gamma(1+\hbar)\Gamma(1+5\hbar)} & \displaystyle\frac{2^{2\hbar}\Gamma(\frac{1}{2}-2\hbar)\Gamma^2(\frac{1}{2}+\hbar)\Gamma(1+3\hbar)}{\sqrt{3}\pi\Gamma(\frac{1}{2}-\hbar)\Gamma(1+4\hbar)} & \csc(3\pi\hbar)\sin(\pi\hbar)
    \end{pmatrix}.
\end{multline}
In order to verify that this expression is equal to the quantum \(6j\)-symbols, we use the following formulas:
\begin{equation}
    \Gamma(z)\Gamma(z+\frac{1}{2})=2^{1-2z}\sqrt{\pi}\,\Gamma(2z)\,,
\end{equation}
\begin{equation}
    \log \Gamma(1+z)=-\gamma z-\frac{1}{2}\log \frac{\sin(\pi z)}{\pi z}-\sum_{k=1}^{\infty}\frac{\zeta(2k+1)}{2k+1}z^{2k+1}\,.
\end{equation}
Applying these relations for Gamma-functions in~\eqref{UPhi-[2]^3} and eliminating odd zeta values, one obtains:
\begin{equation}
    U\cdot \Phi^\KZ_{\mathbb{Q}} = U_q = \begin{pmatrix}
        \frac{1}{[6]} & - \frac{1}{[2]}\sqrt{\frac{[5]}{[3]}} & \frac{\sqrt{[5]}}{[3]} \\
        \frac{1}{[2]}\sqrt{\frac{[5]}{[3]}} & -\frac{1}{[2]} & -\frac{1}{\sqrt{[3]}} \\
        \frac{\sqrt{[5]}}{[3]} & \frac{1}{\sqrt{[3]}} & \frac{1}{[3]}
    \end{pmatrix}.
\end{equation}

\subsection{Perturbative checks}
\label{subsec:perturbative-check}
To verify our relation for other representations, we did not solve the Knizhnik--Zamolodchikov equation; instead, we used the already known series for the associator (\ref{Drseries}) and checked the formula up to some order in $h$. We present a list of all our checks:
\begin{itemize}
    \item \([1]\otimes[1]\otimes[1]\) for \(\mathfrak{sl}_N\) up to degree 12;
    
    \item \([2]\otimes[1]\otimes[1]\) for \(\mathfrak{sl}_N\) up to degree 12;
    
    \item \([2]\otimes[2]\otimes[1]\) for \(\mathfrak{sl}_N\) up to degree 12;
    
    \item \([2]\otimes[2]\otimes[2]\) for \(\mathfrak{sl}_N\) up to degree 8;
    
    \item \([2,1]\otimes[2,1]\otimes[2,1]\) for \(\mathfrak{sl}_N\):
    \begin{itemize}
        \item[--] For \([6,2,1]\) up to degree 10;
        \item[--] For \([5, 1, 1, 1, 1]\) up to degree 10;
        \item[--] For \([5, 2, 2]\) up to degree 7;
        \item[--] For \([4, 4, 1]\) up to degree 7;
        \item[--] For \([5,3,1]\) up to degree 7;
        \item[--] For \([5, 2, 1, 1]\) up to degree 7.
    \end{itemize}

    \item \([2,2]\otimes[2,2]\otimes[2,2]\) for \(\mathfrak{sl}_N\):
    \begin{itemize}
        \item[--] For \([5, 3, 2, 1, 1]\) up to degree 10;
        \item[--] For \([5, 4, 2, 1]\) up to degree 7;
        \item[--] For \([4, 4, 2, 2]\) up to degree 8.
    \end{itemize}

    \item \([1]\otimes[1]\otimes[1]=[3]\oplus2\cdot[2,1]\oplus [1,1,1]\oplus 3\cdot[1]\) for \(\mathfrak{so}_5\):
    \begin{itemize}
        \item[--] For \([2,1]\) up to degree 12.
        \item[--] For \([1]\) up to degree 8.
    \end{itemize}
\end{itemize}
The explicit matrices \(A\), \(B\), and the \(6j\)-symbol matrices are written in Appendix~\ref{sec:C}.\\

\bigskip

Let us note several important points obtained in the process of these checks:
\begin{enumerate}
    \item For the checks of representations of the algebra \(\mathfrak{sl}_N\), we used all the matrices for \(\mathfrak{sl}_2\). Why can we do this? First, we are interested not in the full \(6j\) symbol matrices, but in the matrices on the multiplicity space (the space of highest weight vectors); for \(\mathfrak{sl}_N\), the \(6j\) symbols do not depend on \(N\). Second, the eigenvalues of the split Casimir operator are expressed in terms of the eigenvalues of the quadratic Casimir:
    \begin{equation}
        \lambda_{\mu} = \frac{1}{2}(C(\mu)-C(\lambda_1)-C(\lambda_2)),\quad \mu \in \lambda_1\otimes\lambda_2\,.
        \label{eq:quadratic-casimir}
    \end{equation}
In turn, for the quadratic Casimir eigenvalue in the algebra \(\mathfrak{sl}_N\), there is the following formula:
\begin{equation}
    C(\mu) = \sum_{(i,j)\in \mu}(j-i) + \frac{N|\mu|}{2}-\frac{|\mu|^2}{2N}
    \label{eq:casimir-sl_N}
\end{equation}
where the sum goes over the boxes of the Young diagram \(\mu\).

The constant part involving \(N\) and $C(\lambda_1)$, $C(\lambda_2)$ does not contribute to the associator, since equation~\eqref{eq:eq-on-Phi} determining it is invariant under scalar shifts of $A$ and $B$.


    \item In our examples, all terms containing multiple zeta values canceled, and it was sufficient to simply remove the odd zeta values.

    \item Furthermore, the examples involving the cube of the representations \([2,1]\) and $[2,2]$ are interesting, since multiplicities arise there. As is well known, when they appear, a certain freedom arises in the choice of the \(6j\)-symbols. Namely, one may act by a rotation matrix in the multiplicity space (which may itself contain \(\hbar\)). And, strictly speaking, the formula \(U_q = U \Phi\) ceases to be correct, but we have found the appropriate rotation matrix (up to some order) in order to fix the formula.

\end{enumerate}

\section{Conclusion}
\label{sec::4}
In this paper, we have studied the relation between associators and \(6j\)-symbols, which had not been explicitly investigated before. A possible exact formula relating them, \(U_q= U\cdot\Phi^{\KZ}_{\mathbb{Q}}\) , was proposed but not proved, supported by many examples.\\
Moreover, this formula can possibly be reformulated in terms of finding the appropriate twist from formula (\ref{tw}), but a search in this direction has not yet been carried out. One can further conjecture that the associator \(\Phi^{\KZ}_{\mathbb{Q}}\) is a rational associator for which no explicit construction existed previously (apart from the iterative solution of the pentagon and hexagon equations by Bar-Natan), and one can consider the quasi-Hopf algebra \((\mathcal{U}(\mathfrak{g})[[h]], \Delta, \mathcal{R}^{\KZ}, \Phi^{\KZ}_{\mathbb{Q}})\). This, for example, explains why, when computing knot invariants by means of the Kontsevich integral and the associator \(\Phi^{\KZ}\), transcendental zeta values do not contribute to the resulting polynomial.

The open problems are to prove this formula, or find the appropriate twist, or verify more examples for different algebras and larger representations.

\section*{Acknowledgments}

The work was partially funded within the state assignment of the Institute for Information Transmission Problems of RAS.

\appendix

\section{Example of a twist}
\label{sec:A}
Let us demonstrate, using the example of $\mathfrak{sl}_2$, how one can obtain a deformation-quantized algebra from the Drinfeld-quantized algebra by means of a twist in the first order. Note that to first order, both algebras have a trivial associator, but different $\CR$-matrices and comultiplications.

We consider the algebra $\sl_2$ with the standard comultiplication $\Delta=\mathds{1}\otimes a + a\otimes \mathds{1}$:
\begin{equation}
    [H,E]=2E\,,\quad
    [F,H]=2F\,,\quad
    [E,F]=H\,.
\end{equation}
We aim to obtain the following comultiplication:
\begin{equation}
    \Delta_q(E) = E \otimes \mathds{1} + e^{\frac{h}{2}H} \otimes E = E \otimes \mathds{1} + \mathds{1} \otimes E + \frac{h}{2} H \otimes E + \mathcal{O}(h^2)\,,
\end{equation}
\begin{equation}
    \Delta_q(F) = F \otimes e^{-\frac{h}{2}H} + \mathds{1} \otimes F = \mathds{1} \otimes F + F \otimes \mathds{1} - \frac{h}{2} F \otimes H + \mathcal{O}(h^2)\,,
\end{equation}
\begin{equation}
    \Delta_q(H) = H \otimes \mathds{1} + \mathds{1} \otimes H\,.
\end{equation}
Take the following twist:
\begin{equation}
    \mathcal{F} = \mathds{1} \otimes \mathds{1} -\frac{h}{2}F\otimes E + \mathcal{O}(h^2)\,.
\end{equation}
Then, after the twist we obtain:
\begin{equation}
   \Delta_q(a)= \Delta^{\mathcal{F}}(a) = \mathcal{F}\Delta(a)\mathcal{F}^{-1} = \Delta(a) - \frac{h}{2}[F \otimes E, \Delta(a)]\,.
\end{equation}
One can verify that we obtain the desired comultiplications.

We next find the new ${\cal R}^{\mathcal{F}}$-matrix (where ${\cal R} = \exp\left(\frac{h}{2}\tau\right) = 1 \otimes 1 + \frac{h}{2}\tau + \cdots$):
\begin{equation}
    {\cal R}^{\mathcal{F}} = \mathcal{F}_{21}{\cal R}\mathcal{F}^{-1}=\mathds{1}\otimes \mathds{1}+\frac{h}{4} H\otimes H + h F \otimes E\,,
\end{equation}
\begin{equation}
    {\cal R}^{\mathcal{F}}= \begin{pmatrix}
        1+\frac{h}{4} & & & \\
        & 1- \frac{h}{4} & 0 & \\
        & h & 1- \frac{h}{4} &\\
        & & & 1 + \frac{h}{4}
    \end{pmatrix}.
\end{equation}
This coincides with the standard ${\cal R}$-matrix for $U_q(\sl_2)$ with $q = e^{\frac{h}{2}}$:
\begin{equation}
    \mathcal{R}= \begin{pmatrix}
        q^{\frac{1}{2}} & & & \\
        & q^{-\frac{1}{2}} & 0 & \\
        &q^{\frac{1}{2}} - q^{-\frac{3}{2}}& q^{-\frac{1}{2}}& \\ & & & q^{\frac{1}{2}}
    \end{pmatrix}.
\end{equation}


\section{Computation of KZ associators}
\label{sec:B}

\textbf{Note:} Here we perform computation for the $\sl_2$ though Sections~\ref{subsec:exac-fund-sl_N} and~\ref{subsec:exac-sym-sl_N} state that verification holds for the $\sl_N$ case. This extrapolation could be explained with~\eqref{eq:eq-on-Phi},~\eqref{eq:quadratic-casimir} and~\eqref{eq:casimir-sl_N}. Full explanation is presented in Sec.~\ref{subsec:perturbative-check}.

We now present an example of the complete Drinfeld quantization of the Lie algebra $\mathfrak{sl}_2$. That is, we have an algebra with generators $\{E,F,H\}$ and the following comultiplication:
\begin{equation}
      \Delta(E) = E \otimes \mathds{1} + \mathds{1} \otimes E, \quad
\Delta(F) = F \otimes \mathds{1} + \mathds{1} \otimes F, \quad
\Delta(H) = H \otimes \mathds{1} + \mathds{1} \otimes H.
\end{equation}
and with the standard commutation relations:
\begin{equation}
[H,E] = 2E, \qquad [H,F] = -2F, \qquad [E,F] = H.
\end{equation}
We define the ${\cal R}$-matrix as ${\cal R} = \exp(h \tau / 2)$, where the invariant tensor for $\mathfrak{sl}_2$ is $\tau = E\otimes F + F \otimes E + \frac{1}{2} H\otimes H$.

Now we find the Drinfeld associator by solving the Knizhnik--Zamolodchikov equation. 

\paragraph{Case of ${[1]\otimes[1]\otimes[1]}$.}
Now, we work in the fundamental representation. We write down the KZ equation:
\begin{equation}\label{kz}
    \frac{dG(z)}{dz} = \hbar \left( \frac{A}{z} - \frac{B}{1-z} \right) G(z)\,,
\end{equation}
where $A = \tau \otimes \mathds{1}$ and $B=\mathds{1} \otimes \tau$.

We work in the basis $(\Box\otimes\Box)\otimes\Box$, so that $A$ becomes diagonal:
\begin{equation}
    A' = \Lambda, \qquad B' = U^{-1}\cdot \Lambda \cdot U\,,
\end{equation}
where $\Lambda$ is the diagonal matrix of eigenvalues, and $U:([1]\otimes[1])\otimes[1]\rightarrow [1]\otimes([1]\otimes[1])$ is the matrix of $6j$-symbols:
\begin{equation}
  \Lambda = \left(\begin{array}{ c  c  c}
    \frac{1}{2} \cdot\mathbf{1} & & \\
     & \lambda &\\
     & & \lambda
  \end{array}\right), \qquad \lambda = \left(\begin{array}{ c  c }
    \frac{1}{2}   & \\
     & -\frac{3}{2} 
  \end{array}\right)\,,
  \label{eq:lambda}
\end{equation}
\begin{equation}
    U = \left(\begin{array}{ c  c  c}
    \mathbf{1} & & \\
     & u &\\
     & & u
  \end{array}\right), \qquad u =\left(\begin{array}{ c  c }
    -\frac{1}{2}   & \frac{\sqrt{3}}{2} \\
     -\frac{\sqrt{3}}{2}  & -\frac{1}{2} 
  \end{array}\right)\,.
\end{equation}
In this basis, the matrix $B$ also becomes block-diagonal:
\begin{equation}
    B' = \left(\begin{array}{ c  c  c}
    \frac{1}{2}\cdot\mathbf{1} & & \\
     & b &\\
     & & b
  \end{array}\right), \qquad b = \left(\begin{array}{ c  c }
    -1   & -\frac{\sqrt{3}}{2} \\
     -\frac{\sqrt{3}}{2} & 0
  \end{array}\right)\,.
  \label{eq:B}
\end{equation}
We shall seek the associator in the form:
\begin{equation}
    \tilde \Phi(z) = (1-z)^{-\hbar B} G(z) z^{-\hbar A}\,.
    \label{eq:Phi-def}
\end{equation}

\begin{equation}
    \frac{d \tilde \Phi}{dz} = \frac{\hbar}{z}(1-z)^{-\hbar B} A (1-z)^{\hbar B}\tilde \Phi - \frac{\hbar}{z}\tilde\Phi A,\qquad\text{with } \tilde\Phi(0)=\mathds{1}\,.
\end{equation}
From the imposed initial condition $\tilde\Phi(0)=\mathds{1}$ and form of matrices \eqref{eq:lambda}--\eqref{eq:B}, we can search $\tilde\Phi(z)$ as follows:
\begin{equation}
    \tilde\Phi(z) = \left[\begin{array}{ c  c  c}
    \mathbf{1} & & \\
     & \phi(z) &\\
     & & \phi(z)
  \end{array}\right], \qquad \phi(z) = \left(\begin{array}{ c  c }
    \phi_1(z)   & \phi_3(z) \\
     \phi_2(z) & \phi_4(z)
  \end{array}\right)\,.
  \label{eq:Phi}
\end{equation}
Performing the following change:
\begin{equation}
    \left(\begin{array}{ c  c }
    \phi_1   & \phi_3 \\
     \phi_2 & \phi_4
  \end{array}\right) = U^{-1} F=
  \left(\begin{array}{ c  c }
    f_1+\sqrt{3}f_2   & f_3+\sqrt{3}f_4 \\
     -\sqrt{3}f_1+f_2 & -\sqrt{3}f_2+f_4
  \end{array}\right)\,.
  \label{eq:phi-change}
\end{equation}
We obtain the system of differential equations on $f_1,f_2,f_3,f_4$:
\begin{equation}
    \begin{cases}
    z\displaystyle\frac{df_1}{dz} = -\frac{3}{2}\hbar f_1 + \frac{\sqrt{3}}{2}\hbar (1-z)^{-2\hbar}f_2 \\
    z\displaystyle\frac{df_2}{dz} = \frac{\sqrt{3}}{2}\hbar (1-z)^{2\hbar} f_1 - \frac{1}{2}\hbar f_2 \\
    z\displaystyle\frac{df_3}{dz} = \frac{1}{2}\hbar f_3 + \frac{\sqrt{3}}{2}\hbar (1-z)^{-2\hbar}f_4 \\
    z\displaystyle\frac{df_4}{dz} = \frac{\sqrt{3}}{2}\hbar (1-z)^{2\hbar} f_3 + \frac{3}{2}\hbar f_4
    \end{cases},\qquad
    \begin{cases}
        f_1(0)=\displaystyle\frac{1}{4} \\
        f_2(0)=\displaystyle\frac{\sqrt{3}}{4} \\
        f_3(0)=-\displaystyle\frac{\sqrt{3}}{4} \\
        f_4(0)=\displaystyle\frac{1}{4}
    \end{cases}.
\end{equation}
This system can be rewritten regarding Gauss hypergeometric equation:
\begin{equation}
    \begin{cases}
        z(1-z)f_1'' + (1+2\hbar+(-4\hbar-1)z)f_1' - 3\hbar^2 f_1 = 0 \\
        z(1-z)f_2'' + (1+2\hbar-z)f_2' + \hbar^2 f_2 = 0 \\
        z(1-z)f_3'' + (1-2\hbar-z)f_3' + \hbar^2 f_3 = 0 \\
        z(1-z)f_4'' + (1-2\hbar+(4\hbar-1)z)f_4' - 3\hbar^2 f_4 = 0
    \end{cases}
\end{equation}
In order to obtain solutions we additionally need conditions for first derivatives:
\begin{equation}
    \begin{cases}
    f_1'(0)=-\frac{3}{2}\hbar f_1'(0) + \sqrt{3}\hbar^2 f_2(0) + \frac{\sqrt{3}}{2}\hbar f_2'(0) \\
    f_2'(0)=-\sqrt{3}\hbar^2 f_1(0) + \frac{\sqrt{3}}{2}\hbar f_1'(0) - \frac{1}{2}\hbar f_2'(0) \\
    f_3'(0)=\frac{1}{2}\hbar f_3'(0) + \sqrt{3}\hbar^2 f_4(0) + \frac{\sqrt{3}}{2}\hbar f_4'(0) \\
    f_4'(0)=-\sqrt{3}\hbar^2 f_3(0) + \frac{\sqrt{3}}{2}\hbar f_3'(0) + \frac{3}{2}\hbar f_4'(0) 
\end{cases} \quad\Rightarrow\quad
\begin{cases}
    f_1'(0)=\frac{3\hbar^2}{4(1+2\hbar)} \\
    f_2'(0)=-\frac{\sqrt{3}\hbar^2}{4(1+2\hbar)}\\
    f_3'(0)=\frac{\sqrt{3}\hbar^2}{4(1-2\hbar)} \\
    f_4'(0)=\frac{3\hbar^2}{4(1-2\hbar)}
\end{cases}
\end{equation}
Thus, the solution is:
\begin{equation}
    \begin{cases}
        f_1=\frac{1}{4}{}_2F_1(\hbar,3\hbar|1+2\hbar|z)\\
        f_2=\frac{\sqrt{3}}{4}{}_2F_1(\hbar,-\hbar|1+2\hbar|z)\\
        f_3=-\frac{\sqrt{3}}{4}{}_2F_1(\hbar,-\hbar|1-2\hbar|z)\\
        f_4=\frac{1}{4}{}_2F_1(-\hbar,-3\hbar|1-2\hbar|z)\\
    \end{cases}
    \label{eq:f-ans}
\end{equation}
Implementing \eqref{eq:f-ans} in \eqref{eq:phi-change} gives us the answer for associator:
\begin{equation}
    \Phi= \frac{1}{4}\left(\begin{array}{ c  c }
    \displaystyle\frac{\Gamma(1+2\hbar)\Gamma(1-2\hbar)}{\Gamma(1+\hbar)\Gamma(1-\hbar)}+3\displaystyle\frac{\Gamma^2(1+2\hbar)}{\Gamma(1+3\hbar)\Gamma(1+\hbar)}   & -\sqrt{3}\left(\displaystyle\frac{\Gamma^2(1-2\hbar)}{\Gamma(1-3\hbar)\Gamma(1-\hbar)}-\frac{\Gamma(1-2\hbar)\Gamma(1+2\hbar)}{\Gamma(1+\hbar)\Gamma(1-\hbar)}\right) \\
     -\sqrt{3}\left(\displaystyle\frac{\Gamma(1+2\hbar)\Gamma(1-2\hbar)}{\Gamma(1+\hbar)\Gamma(1-\hbar)}-\displaystyle\frac{\Gamma^2(1+2\hbar)}{\Gamma(1+3\hbar)\Gamma(1+\hbar)}\right) & 3\displaystyle\frac{\Gamma^2(1-2\hbar)}{\Gamma(1-3\hbar)\Gamma(1-\hbar)}+\frac{\Gamma(1-2\hbar)\Gamma(1+2\hbar)}{\Gamma(1+\hbar)\Gamma(1-\hbar)}
  \end{array}\right).
\end{equation}

\paragraph{Case of $[2]\otimes[2]\otimes[2]$}
Similarly, one can solve the equation for the tensor cube of the symmetric representation:
\begin{equation}
    ([2] \otimes [2]) \otimes [2] = ([4] \oplus [2] \oplus [0]) \otimes [2] = [6] \oplus 2 \cdot [4] \oplus 3 \cdot [2] \oplus [0]\,.    \label{eq:sl2-2-decomp}
\end{equation}
We perform computation for $3\times3$ block corresponding to $3\cdot[2]$ in decomposition and we have following matrices:
\begin{equation}
    \Lambda = \begin{pmatrix}
        2 & 0 & 0\\
        0 & -2 & 0\\
        0 & 0 & -4
    \end{pmatrix},\quad B = \begin{pmatrix}
        -3 & -\sqrt{\frac{5}{3}} & 0\\
        -\sqrt{\frac{5}{3}} & -1 & -\frac{4}{\sqrt{3}}\\
        0 & -\frac{4}{\sqrt{3}} & 0
    \end{pmatrix}, \quad U_q = \begin{pmatrix}
        \frac{1}{[6]} & - \frac{1}{[2]}\sqrt{\frac{[5]}{[3]}} & \frac{\sqrt{[5]}}{[3]} \\
        \frac{1}{[2]}\sqrt{\frac{[5]}{[3]}} & -\frac{1}{[2]} & -\frac{1}{\sqrt{[3]}} \\
        \frac{\sqrt{[5]}}{[3]} & \frac{1}{\sqrt{[3]}} & \frac{1}{[3]}
    \end{pmatrix}.
\end{equation}
As before, we have the equation:
\begin{equation}
    \frac{d\tilde \Phi}{dz} = \frac{\hbar}{z}(1-z)^{-\hbar B} A (1-z)^{\hbar B}\tilde\Phi - \frac{\hbar}{z}\tilde\Phi A,\qquad\text{with } \tilde\Phi(0)=\mathds{1}\,.
\end{equation}
We perform the following change:
\begin{equation}
    \tilde\Phi = U^{-1}F,\qquad F=||f_{ij}(z)||_{1\le i,j\le3}\,.
\end{equation}
Let us carry out the computations only for the first column of the matrix $F=||f_{ij}(z)||_{1\le i,j\le3}$; for the remaining ones, everything is done analogously. Thus, we have the following system of equations:
\begin{equation}
    \begin{cases}
    \displaystyle \frac{d f_{11}}{dz} = \frac{\hbar}{z} \left( -5 f_{11} + \sqrt{\frac{5}{3}}(1-z)^{-4\hbar}f_{21}\right) \\
    \displaystyle \frac{d f_{21}}{dz} = \frac{\hbar}{z} \left( \sqrt{\frac{5}{3}} (1-z)^{4\hbar} f_{11} - 3 f_{21} + \frac{4}{\sqrt{3}}(1-z)^{-2\hbar} f_{31}\right) \\
    \displaystyle \frac{d f_{31}}{dz} = \frac{\hbar}{z} \left( \frac{4}{\sqrt{3}}(1-z)^{2\hbar} f_{21} - 2 f_{31}\right)
    \end{cases},\qquad
    \begin{cases}
        f_{11}(0)=\displaystyle\frac{1}{6} \\
        f_{21}(0)=\displaystyle\frac{1}{2}\sqrt{\frac{5}{3}} \\
        f_{31}(0)=\displaystyle\frac{\sqrt{5}}{3} 
    \end{cases}.
    \label{eq:sl2-2-first-col}
\end{equation}
From this system, one can derive an equation for \(f_{31}\):
\begin{equation}
    z^2(1-z)^2 f'''_{31} + z (1-z) (3 + 10\hbar - (3 + 2\hbar)z) f''_{31} + (1 + 10 \hbar + 24 \hbar^2 - (2 + 10 \hbar + 4 \hbar^2) z + (1 + 2 \hbar - 8 \hbar^2) z^2) f'_{31} + 4 \hbar^2 (1 + 6 \hbar) f_{31}= 0\,.
\end{equation}
We expect that its solutions will be hypergeometric functions, namely \({}_3F_2\). But for this, one needs to carefully bring the equations into the required form. To do so, let us make the following change of variables:
\begin{equation}
    x = z + \frac{1}{z}\,,\qquad f_{31}(z) = z^{-2\hbar} g_{31}\left( -\frac{4}{-(z+\frac{1}{z})+2} \right)\,.
\end{equation}
After this, we obtain a hypergeometric equation:
\begin{equation}
    x^2(1 - x) g'''_{31}(x) + \frac{x}{2} \bigl(6 + 8 \hbar - (9 + 8 \hbar) x\bigr) g''_{31}(x) + g'_{31}(x) + \frac{x}{2} \bigl(2 + 8 \hbar - 8 \hbar^2 - (6 + 13 \hbar + 6 \hbar^2) x\bigr) - 16 \hbar^3 g_{31}(x) = 0\,,
\end{equation}
\begin{equation}
    g_{31}(x) = \frac{(-1)^{-2\hbar} 16^{-\hbar}\sqrt{5}}{3} x^{2\hbar} {}_3 F_2\left( \frac{1}{2} - \hbar, -2\hbar, 1+\hbar \bigg| 1-4\hbar, 1+2\hbar \bigg| x \right)\,.
\end{equation}
In order to find the value of our solution at \(z = 1\), let us use the asymptotic expansion formula for the hypergeometric function for $z\rightarrow\infty$:
\begin{multline}
    {}_3 F_2\left( a_1,a_2,a_3 \bigg| b_1,b_2 \bigg| z \right) = \frac{\Gamma(b_1)\Gamma(b_2)}{\Gamma(a_1)\Gamma(a_2)\Gamma(a_3)}\cdot \\
    \cdot \Biggl(\frac{\Gamma(a_1)\Gamma(a_2-a_1)\Gamma(a_3-a_1)}{\Gamma(b_1-a_1)\Gamma(b_2-a_1)}(-z)^{-a_1}{}_3 F_2\left( a_1,a_1-b_1+1,a_1-b_2+1 \bigg| a_1-a_2+1,a_1-a_3+1 \bigg| \frac{1}{z} \right) + \\
    \frac{\Gamma(a_2)\Gamma(a_1-a_2)\Gamma(a_3-a_2)}{\Gamma(b_1-a_2)\Gamma(b_2-a_2)}(-z)^{-a_2}{}_3 F_2\left( a_2,a_2-b_1+1,a_2-b_2+1 \bigg| -a_1+a_2+1,a_2-a_3+1 \bigg| \frac{1}{z} \right) + \\
    \frac{\Gamma(a_3)\Gamma(a_1-a_3)\Gamma(a_2-a_3)}{\Gamma(b_1-a_3)\Gamma(b_2-a_3)}(-z)^{-a_3}{}_3 F_2\left( a_3,a_3-b_1+1,a_3-b_2+1 \bigg| -a_1-a_3+1,-a_2+a_3+1 \bigg| \frac{1}{z} \right)
    \Biggr)\,.
\end{multline}
This yields the answer:

\begin{multline}
    F 
    =\begin{pmatrix}
        \displaystyle\frac{\sec(2\pi\hbar)}{2+4\cos(2\pi\hbar)} & \displaystyle\frac{2^{-6\hbar}\Gamma(1-3\hbar)\Gamma^2(\frac{1}{2}-2\hbar)\Gamma(2\hbar)}{\sqrt{15\pi}\Gamma(\frac{1}{2}-\hbar)\Gamma(-5\hbar)\Gamma(1+\hbar)} & \displaystyle\frac{\sqrt{5}\Gamma(1-3\hbar)\Gamma(-3\hbar)}{\Gamma(1-5\hbar)\Gamma(-\hbar)}\\
        \displaystyle\frac{2^{4\hbar}\sqrt{15}\hbar\Gamma(\frac{1}{2}-\hbar)\Gamma(3\hbar)\Gamma^2(\frac{1}{2}+2\hbar)}{2\pi\Gamma(\frac{1}{2}+\hbar)\Gamma(1+5\hbar)} & \displaystyle\frac{\sec(2\pi\hbar)-2}{2} & \displaystyle-\frac{2^{-2\hbar}\Gamma(1-3\hbar)\Gamma^2(\frac{1}{2}-\hbar)\Gamma(\frac{1}{2}+2\hbar)}{\sqrt{3}\pi\Gamma(1-4\hbar)\Gamma(\frac{1}{2}+\hbar)}\\
        \displaystyle\frac{\sqrt{5}\Gamma^2(1+3\hbar)}{3\Gamma(1+\hbar)\Gamma(1+5\hbar)} & \displaystyle\frac{2^{2\hbar}\Gamma(\frac{1}{2}-2\hbar)\Gamma^2(\frac{1}{2}+\hbar)\Gamma(1+3\hbar)}{\sqrt{3}\pi\Gamma(\frac{1}{2}-\hbar)\Gamma(1+4\hbar)} & \csc(3\pi\hbar)\sin(\pi\hbar)
    \end{pmatrix}.
\end{multline}


\section{Perturbative checks}
\label{sec:C}
In this section, we shall verify our formula perturbatively, using the series (\ref{Drseries}), discarding all irrational zeta values. The series for the associator was taken from the website of S.~Duzhin; it is available at \url{https://www.pdmi.ras.ru/~arnsem/dataprog/associator/}.\\
\begin{equation}
\begin{aligned}
        \log\Phi^{KZ} &= -\hbar^2 \zeta_2 C_{11} - \hbar^3 \zeta_3 (C_{12}+C_{21})-\\
        &-\hbar^4\frac{2}{5}\zeta_2^2(C_{13}+C_{31}) - \hbar^4 \frac{1}{10}\zeta_2^2 C_{22}- \\
        &- \hbar^5 \zeta_5 (C_{14}+C_{41}) + \hbar^5 (\zeta_2\zeta_3 - 2\zeta_5)(C_{23}+C_{32}) + \hbar^5 \frac{\zeta_2\zeta_3 - \zeta_5}{2}[C_{11},C_{12}] + \hbar^5 \frac{\zeta_2\zeta_3 - 3\zeta_5}{2}[C_{11},C_{21}]- \\
        &- \hbar^6 \frac{8}{35} \zeta_2^3 (C_{15}+C_{51}) + \hbar^6 \left(\frac{1}{2}\zeta_3^2-\frac{6}{35}\zeta_2^3\right)(C_{24}+C_{42}) + \hbar^6 \left(\zeta_3^2-\frac{23}{70}\zeta_2^3\right)C_{33}+ \hbar^6 \left(-\frac{19}{105}\zeta_2^3+\zeta_3^2\right)[C_{11},C_{13}] + \\
        &+ \hbar^6 \left(-\frac{69}{140}\zeta_2^3+\frac{3}{2}\zeta_3^2\right)[C_{11},C_{22}]+ \hbar^6 \left(-\frac{17}{105}\zeta_2^3\right)[C_{11},C_{31}] + \hbar^6 \left(\frac{2}{105}\zeta_2^3-\frac{1}{2}\zeta_3^2\right)[C_{12},C_{21}]+\dots
\end{aligned} 
\end{equation}
where
\begin{equation}
    \zeta_n = \zeta(n),\quad C_{kl} = \text{ad}_B^{k-1}\text{ad}_A^{l-1}[A,B]\,.
\end{equation}
In what follows, for each case, we shall specify the matrices $A=\Lambda$ and $B=U^{-1}\Lambda U$ used in the associator and the matrix of $6j$ symbols for which we verify the formula.\\
Note that here:
\begin{equation}
    [n]_q = \frac{q^{n}-q^{-n}}{q - q^{-1}}\,.
\end{equation}

\begin{itemize}
    \item $[1]\otimes [1] \otimes [1]  $  for $\mathfrak{sl}_N$:
    \begin{equation}
        \Lambda=  \left(\begin{array}{ c  c }
    \frac{1}{2}   & \\
     & -\frac{3}{2} 
  \end{array}\right)\,, \quad U_q=\left(\begin{array}{ c  c }
    -\frac{1}{[2]}   & \frac{\sqrt{[3]}}{[2]} \\
     -\frac{\sqrt{[3]}}{[2]}  & -\frac{1}{[2]} 
  \end{array}\right)\,.
    \end{equation}

    \item $[2]\otimes [1] \otimes [1]  $  for $\mathfrak{sl}_N$ : 
    \begin{equation}
    \Lambda = \begin{pmatrix}
        1 & 0 \\
        0 & -2
    \end{pmatrix},\quad
    U_q = \begin{pmatrix}
        -\frac{1}{\sqrt{[3]}} & \sqrt{\frac{[4]}{[2][3]}} \\
        -\sqrt{\frac{[4]}{[2][3]}} & -\frac{1}{\sqrt{[3]}}
    \end{pmatrix}.
\end{equation}

    \item  $[2]\otimes [2] \otimes [1]  $  for $\mathfrak{sl}_N$:
    \begin{equation}
    \Lambda = \begin{pmatrix}
        1 & 0 \\
        0 & -2
    \end{pmatrix},\quad
    U_q = \begin{pmatrix}
        -\frac{1}{\sqrt{[3]}} & \sqrt{\frac{[4]}{[2][3]}} \\
        -\sqrt{\frac{[4]}{[2][3]}} & -\frac{1}{\sqrt{[3]}}
    \end{pmatrix}.
\end{equation}

    \item  $[2]\otimes [2] \otimes [2]  $  for $\mathfrak{sl}_N$:
    \begin{equation}
        \Lambda = \begin{pmatrix}
        2 & 0 & 0\\
        0 & -2 & 0\\
        0 & 0 & -4
    \end{pmatrix},\quad  U_q = \begin{pmatrix}
        \frac{1}{[6]} & - \frac{1}{[2]}\sqrt{\frac{[5]}{[3]}} & \frac{\sqrt{[5]}}{[3]} \\
        \frac{1}{[2]}\sqrt{\frac{[5]}{[3]}} & -\frac{1}{[2]} & -\frac{1}{\sqrt{[3]}} \\
        \frac{\sqrt{[5]}}{[3]} & \frac{1}{\sqrt{[3]}} & \frac{1}{[3]}
    \end{pmatrix}.
    \end{equation}

    \item $[2,1]\otimes [2,1] \otimes [2,1]  $  for $\mathfrak{sl}_N$ :

    \begin{equation}
                \Lambda_{[6, 2, 1]} = \begin{pmatrix}
                5 & 0 \\
                0 & 3
            \end{pmatrix},\quad U_q^{[6, 2, 1]} :=
        \begin{pmatrix}
        \dfrac{q}{1 + q^2} & \dfrac{\sqrt{(q^2 + q + 1)(q^2 - q + 1)}}{1 + q^2} \\[12pt]
        \dfrac{\sqrt{(q^2 + q + 1)(q^2 - q + 1)}}{1 + q^2} & -\dfrac{q}{1 + q^2}
        \end{pmatrix},
    \end{equation}

    \begin{equation}
        \Lambda_{[5,4]} = \begin{pmatrix}
        5 & 0 \\
        0 & 3
    \end{pmatrix},\quad
      U_q^{[5, 4]} :=
        \begin{pmatrix}
        \dfrac{q}{1 + q^2} & \dfrac{\sqrt{(q^2 + q + 1)(q^2 - q + 1)}}{1 + q^2} \\[12pt]
        \dfrac{\sqrt{(q^2 + q + 1)(q^2 - q + 1)}}{1 + q^2} & -\dfrac{q}{1 + q^2}
        \end{pmatrix},
    \end{equation}

    \begin{equation}
        \Lambda_{[5,1,1,1,1]} = \begin{pmatrix}
        3 & 0 \\
        0 & -3
    \end{pmatrix},\quad U_q^{[5, 1, 1, 1, 1]} :=
\begin{pmatrix}
\dfrac{1}{q^3 + q^{-3}} & \dfrac{\sqrt{q^6 + 1 + q^{-6}}}{q^3 + q^{-3}} \\[10pt]
\dfrac{\sqrt{q^6 + 1 + q^{-6}}}{q^3 + q^{-3}} & -\dfrac{1}{q^3 + q^{-3}}
\end{pmatrix}.
    \end{equation}

\item $[1]\otimes [1] \otimes [1]  $ for $\mathfrak{so}_5$:
    \begin{equation}
    \Lambda_{[2,1]} = \begin{pmatrix}
        1 & 0 \\
        0 & -1 \\
    \end{pmatrix},\quad
    U_q^{[2,1]} = \begin{pmatrix}
        \frac{1}{[2]} & \frac{\sqrt{[3]}}{[2]} \\
        \frac{\sqrt{[3]}}{[2]} & -\frac{1}{[2]}
    \end{pmatrix},
\end{equation}

\begin{equation}
    \Lambda_{[1]}= \begin{pmatrix}
        1 & 0 & 0\\
        0 & -1 & 0\\
        0 & 0 & -4
    \end{pmatrix},\quad
    U^{[1]}_q = \begin{pmatrix}
        -\frac{[3/2]}{[2][5/2]} & \frac{1}{[2]}\sqrt{\frac{[3/2][7/2]}{[5/2][3]}} & -\frac{1}{[5/2]}\sqrt{\frac{[3/2][7/2][4]}{[2][3]}}\\
        
          \frac{1}{[2]}\sqrt{\frac{[3/2][7/2]}{[5/2][3]}} & -\frac{[3/2][5]}{[2][5/2][3]} & -\frac{[3/2]}{[3]}\sqrt{\frac{[4]}{[2][5/2]}} \\
          -\frac{1}{[5/2]}\sqrt{\frac{[3/2][7/2][4]}{[2][3]}} &  -\frac{[3/2]}{[3]}\sqrt{\frac{[4]}{[2][5/2]}} & -\frac{[3/2]}{[5/2][3]}
    \end{pmatrix}.
\end{equation}
    
\end{itemize}
Owing to the cumbersomeness of the expressions, we omit the remaining matrices of quantum \(6j\) symbols; the interested reader can find them at \url{http://wwwth.mpgnet.ru/knotebook/}.

\newpage






\printbibliography

@incollection{drinfeld1989quasi,
  title={Quasi-Hopf algebras and Knizhnik-Zamolodchikov equations},
  author={Drinfeld, V.},
  booktitle={Problems of Modern Quantum Field Theory: Invited Lectures of the Spring School held in Alushta USSR, April 24--May 5, 1989},
  pages={1--13},
  year={1989},
  publisher={Springer}
}

@book{lemur,
  title={The Kontsevich integral for the Kauffman polynomial},
  author={T. Q. T. Le and J. Murakami},
  year={1996},
  journal={Nagoya Math. J }
}

@book{barnatan,
  title={Non-associative tangles},
  author={D. Bar-Natan},
  year={1996},
  journal={Geometric topology},
   pages={139-183},
}

@article{drinfeld1990,
  title={Quasi-Hopf algebras},
  author={V. Drinfeld},
  year={1990},
  journal={Leningrad Math. J.}
}

@article{drinfeld1991,
  title={On quasitriangular quasi-Hopf algebras and a group closely connected with $\mathrm{Gal}(\overline{\mathbb{Q}}/\mathbb{Q})$},
  author={V. Drinfeld},
  year={1991},
  journal={Leningrad Math. J.}
}

@article{alekseevtorossian2012,
  title={The Kashiwara--Vergne conjecture and Drinfeld's associators},
  author={A. Alekseev and C. Torossian},
  year={2012},
  journal={Ann. of Math.}
}

@article{aet2010,
  title={Drinfeld associators, braid groups and explicit solutions of the Kashiwara--Vergne equations},
  author={A. Alekseev and B. Enriquez and C. Torossian},
  year={2010},
  journal={Publ. Math. IH\'ES}
}

@article{anxz2018,
  title={Chern--Simons, Wess--Zumino and other cocycles from Kashiwara--Vergne and associators},
  author={A. Alekseev and F. Naef and X. Xu and C. Zhu},
  year={2018},
  journal={Lett. Math. Phys.}
}

@article{broadhurstkreimer1995,
  title={Knots and Numbers in $\phi^4$ Theory to 7 Loops and Beyond},
  author={D. J. Broadhurst and D. Kreimer},
  year={1995},
  journal={Int. J. Mod. Phys. C}
}

@article{terasoma2014,
  title={All order $\alpha'$-expansion of superstring trees from the Drinfeld associator},
  author={S. Stieberger and T. Terasoma},
  year={2014},
  journal={Phys. Rev. D}
}

@article{kaderli2020,
  title={A note on the Drinfeld associator for genus-zero superstring amplitudes in twisted de Rham theory},
  author={A. Kaderli},
  year={2020},
  journal={J. Phys. A: Math. Theor.}
}

@misc{broedelmatthesschlotterer2015,
  title={Relations between elliptic multiple zeta values and a special derivation algebra},
  author={J. Broedel and N. Matthes and O. Schlotterer},
  year={2015},
  journal={arXiv:1507.02254}
}

@article{hennlectures2015,
  title={Lectures on differential equations for Feynman integrals},
  author={J. M. Henn},
  year={2015},
  journal={J. Phys. A}
}

@article{racah1942,
  title={Theory of Complex Spectra II},
  author={G. Racah},
  year={1942},
  journal={Phys. Rev.}
}

@article{reshetikhinturaev1991,
  title={Invariants of $3$-manifolds via link polynomials and quantum groups},
  author={N. Reshetikhin and V. Turaev},
  year={1991},
  journal={Invent. Math.}
}

@article{teschnervartanov2014,
  title={$6j$ symbols for the modular double, quantum hyperbolic geometry, and supersymmetric gauge theories},
  author={J. Teschner and G. Vartanov},
  year={2014},
  journal={Lett. Math. Phys.}
}

@article{liuetal2019,
  title={$d$-Dimensional SYK, AdS Loops, and $6j$ Symbols},
  author={J. Liu and E. Perlmutter and V. Rosenhaus and D. Simmons-Duffin},
  year={2019},
  journal={JHEP}
}

@misc{alisauskas2002,
  title={$6j$-symbols for symmetric representations of $SO(n)$ as the double series},
  author={S. Ali\v{s}auskas},
  year={2002},
  journal={arXiv:math-ph/0206044}
}

@article{morozovsleptsov2018,
  title={New symmetries for the $U_q(\mathfrak{sl}_N)$ $6$-$j$ symbols from the Eigenvalue conjecture},
  author={A. Morozov and A. Sleptsov},
  year={2018},
  journal={JETP Lett.}
}

@article{alekseevmorozovsleptsov2019,
  title={Interplay between symmetries of quantum $6$-$j$ symbols and the eigenvalue hypothesis},
  author={V. Alekseev and A. Morozov and A. Sleptsov},
  year={2021},
  journal={Lett. Math. Phys.}
}

@misc{laninasleptsov2022,
  title={Tug-the-hook symmetry for quantum $6j$-symbols},
  author={E. Lanina and A. Sleptsov},
  year={2022},
  journal={arXiv:2210.07874}
}

@article{rivoal2000,
  title={La fonction z\^eta de Riemann prend une infinit\'e de valeurs irrationnelles aux entiers impairs},
  author={T. Rivoal},
  year={2000},
  journal={C. R. Acad. Sci. Paris S\'er. I Math.}
}

@article{zudilin2001,
  title={One of the numbers $\zeta(5)$, $\zeta(7)$, $\zeta(9)$, $\zeta(11)$ is irrational},
  author={W. Zudilin},
  year={2001},
  journal={Russian Math. Surveys}
}

@article{ikz2006,
  title={Derivation and double shuffle relations for multiple zeta values},
  author={K. Ihara and M. Kaneko and D. Zagier},
  year={2006},
  journal={Compositio Math.}
}

@article{racinet2002,
  title={Doubles m\'elanges des polylogarithmes multiples aux racines de l'unit\'e},
  author={G. Racinet},
  year={2002},
  journal={Publ. Math. IH\'ES}
}

@article{smirnov2009,
  title={Notes on Chern-Simons Theory in the Temporal Gauge},
  author={A. Smirnov},
  year={2009},
  journal={hep-th/0910.5011, to appear in the Proceedings of International School of Subnuclear Phys., Erice, Italy}
}

@article{morozovsmirnov2010,
  title={Chern-Simons Theory in the Temporal Gauge and Knot Invariants through the Universal Quantum $R$-Matrix},
  author={A. Morozov and A. Smirnov},
  year={2010},
  journal={Nucl. Phys. B}
}

@book{chmutovduzhinmostovoy2011,
  title={Introduction to Vassiliev Knot Invariants},
  author={S. Chmutov and S. Duzhin and J. Mostovoy},
  year={2011},
  journal={Cambridge University Press, arXiv:1103.5628}
}

@article{CS,
	author = "Chern, S.-S. and Simons, J.",
	title = "{Characteristic forms and geometric invariants}",
	doi = "",
	journal = "Annals Math.",
	volume = "99",
	pages = "48--69",
	year = "1974"
}

@article{Witten,
	title={Quantum field theory and the Jones polynomial},
	author={Witten, E.},
	journal={Communications in mathematical physics},
	volume={121},
	number={3},
	pages={351--399},
	year={1989},
	publisher={Springer}
}

@article{freyd1985new,
	title={A new polynomial invariant of knots and links},
	author={Freyd, P. and Yetter, D. and Hoste, J. and Lickorish, W.B.R. and Millett, K. and Ocneanu, A.},
	journal={Bulletin (new series) of the American mathematical society},
	volume={12},
	number={2},
	pages={239--246},
	year={1985},
	publisher={American Mathematical Society}
}

@article{przytycki1988invariants,
	title={Invariants of links of Conway type},
	author={Przytycki, J.H. and Traczyk, K.P.},
	journal={Kobe Journal of Mathematics},
	volume={4},
	number={2},
	pages={115--139},
	year={1988},
	publisher={理学研究科数学専攻},
	eprint = "1610.06679",
	archivePrefix = "arXiv",
	primaryClass = "math.GT"
}

@article{kauffman1990invariant,
  title={An invariant of regular isotopy},
  author={Kauffman, L.},
  journal={Transactions of the American Mathematical Society},
  volume={318},
  number={2},
  pages={417--471},
  year={1990}
}

@article{reshetikhinturaev1990,
  title={Ribbon graphs and their invariants derived from quantum groups},
  author={N. Reshetikhin and V. G. Turaev},
  year={1990},
  journal={Commun. Math. Phys.},
  volume={127},
  pages={1--26}
}

@article{apery1979,
  title={Irrationalit\'e de $\zeta(2)$ et $\zeta(3)$},
  author={R. Ap\'ery},
  year={1979},
  journal={Ast\'erisque},
  volume={61},
  pages={11--13}
}

@article{moore1989classical,
  title={Classical and quantum conformal field theory},
  author={Moore, G. and Seiberg, N.},
  journal={Communications in Mathematical Physics},
  volume={123},
  number={2},
  pages={177--254},
  year={1989},
  publisher={Springer}
}

@article{knizhnikzamolodchikov1984,
  title={Current Algebra and Wess-Zumino Model in Two Dimensions},
  author={V. G. Knizhnik and A. B. Zamolodchikov},
  year={1984},
  journal={Nucl. Phys. B},
  volume={247},
  pages={83--103}
}

@article{lemurakami1996,
  title={Kontsevich's integral for the Homfly polynomial and relations between values of multiple zeta functions},
  author={T. Q. T. Le and J. Murakami},
  year={1996},
  journal={Topology Appl.},
  volume={62},
  pages={193--206}
}

@article{goncharov2001,
  title={Multiple polylogarithms and mixed Tate motives},
  author={A. B. Goncharov},
  year={2001},
  journal={arXiv:math/0103059}
}

@article{deligneGoncharov2005,
  title={Groupes fondamentaux motiviques de Tate mixte},
  author={P. Deligne and A. B. Goncharov},
  year={2005},
  journal={Ann. Sci. \'Ecole Norm. Sup.},
  volume={38},
  pages={1--56}
}

@article{brown2012,
  title={Mixed Tate Motives over $\mathbb{Z}$},
  author={F. Brown},
  year={2012},
  journal={Ann. of Math.},
  volume={175},
  pages={949--976}
}

@article{itoyama2013eigenvalue,
  title={Eigenvalue hypothesis for Racah matrices and HOMFLY polynomials for 3-strand knots in any symmetric and antisymmetric representations},
  author={Itoyama, Hiroshi and Mironov, A and Morozov, A and Morozov, And},
  journal={International Journal of Modern Physics A},
  volume={28},
  number={03n04},
  pages={1340009},
  year={2013},
  publisher={World Scientific}
}

@article{kontsevich1993,
  title={Vassiliev's Knot Invariants},
  author={M. Kontsevich},
  year={1993},
  journal={Adv. Soviet Math.},
  volume={16},
  pages={137--150}
}

@article{labastidaperez1997,
  title={Kontsevich Integral for Vassiliev Invariants from Chern-Simons Perturbation Theory in the Light-Cone Gauge},
  author={J. M. F. Labastida and E. P\'erez},
  year={1997},
  journal={hep-th/9710176}
}

@article{piunikhin1993,
  title={Combinatorial expression for universal Vassiliev link invariant},
  author={S. Piunikhin},
  year={1993},
  journal={hep-th/9302084}
}

@article{lemurakami1994,
  title={The Universal Vassiliev-Kontsevich Invariant for Framed Oriented Links},
  author={T. Q. T. Le and J. Murakami},
  year={1994},
  journal={hep-th/9401016}
}

@article{hecht1975simple,
  title={A simple class of U (N) Racah coefficients and their application},
  author={Hecht, Karl T},
  journal={Communications in Mathematical Physics},
  volume={41},
  number={2},
  pages={135--156},
  year={1975},
  publisher={Springer}
}

@article{gustafson1987whipple,
  title={A Whipple’s Transformation for Hypergeometric Series in U(n) and Multivariable Hypergeometric Orthogonal Polynomials},
  author={Gustafson, RA},
  journal={SIAM journal on mathematical analysis},
  volume={18},
  number={2},
  pages={495--530},
  year={1987},
  publisher={SIAM}
}

@article{wong1979multiplicity,
  title={On the multiplicity-free Wigner and Racah coefficients of U (n)},
  author={Wong, MKF},
  journal={Journal of Mathematical Physics},
  volume={20},
  number={12},
  pages={2391--2397},
  year={1979},
  publisher={American Institute of Physics}
}

@article{artamonov2025calculation,
  title={Calculation of-Symbols for the Lie Algebra},
  author={Artamonov, Dmitry Vyacheslavovich},
  journal={Siberian Mathematical Journal},
  volume={66},
  number={4},
  pages={875--890},
  year={2025},
  publisher={Springer}
}

@article{blohmann2005reconstruction,
  title={Reconstruction of universal Drinfeld twists from representations},
  author={Blohmann, Christian},
  journal={Journal of mathematical physics},
  volume={46},
  number={5},
  year={2005},
  publisher={AIP Publishing}
}

@article{kulish2006twists,
  title={Twists in and their quantizations},
  author={Kulish, PP and Lyakhovsky, VD and Samsonov, ME},
  journal={Journal of Physics A: Mathematical and General},
  volume={39},
  number={24},
  pages={7669--7692},
  year={2006}
}

\end{document}